\documentclass{vldb_cameraready}
\usepackage{verbatim,alltt,amsmath,amssymb}
\usepackage{graphics,graphicx,color}
\usepackage{algorithmic,algorithm}
\usepackage{listings,color,xcolor}
\usepackage[font=small,labelfont=bf]{caption}
\usepackage{subcaption}
\usepackage{nth}
\usepackage{seqsplit}
\usepackage{enumitem}
\setlist{nosep} 
\usepackage{multirow}
\usepackage{appendix}
\usepackage{balance} 
\usepackage{microtype}
\usepackage[hidelinks]{hyperref}
\hypersetup{
    urlcolor=red
}

\usepackage{listings}
\lstdefinestyle{customc}{
    belowcaptionskip=1\baselineskip,
    breaklines=true,
    frame=single,
    xleftmargin=\parindent,
    language=Python,
    showstringspaces=false,
    basicstyle=\footnotesize\ttfamily,
    keywordstyle=\bfseries\color{green!40!black},
    commentstyle=\itshape\color{purple!40!black},
    identifierstyle=\color{blue},
    stringstyle=\color{orange},
    morekeywords={string,uint64_t,std,vector,pair}
}
\begin{document}
\newtheorem{claim}{Claim}
\newtheorem{defn}{Definition}
\newcommand{\hilight}[1]{\colorbox{yellow}{#1}}
\newcommand{\secref}[1]{\S~\ref{sec:#1}}
\newcommand{\appref}[1]{Appendix~\ref{sec:appendix:#1}}
\newcommand{\figref}[1]{Fig.~\ref{fig:#1}}
\newcommand{\srcref}[1]{Fig.~\ref{fig:#1}}
\newcommand{\clmref}[1]{Claim~\ref{claim:#1}}
\newcommand{\tableref}[1]{Table~\ref{tab:#1}}
\newcommand{\eqnref}[1]{Eq.~\ref{eq:#1}}
\newcommand{\insertfig}[4]{\begin{figure}\centering\includegraphics[width=#1\linewidth]{#2}\caption{#3}\label{fig:#4}\end{figure}}
\newcommand{\insertsrc}[3]{\begin{figure}\centering\lstinputlisting{#1}\caption{#2}\label{fig:#3}\end{figure}}
\newcommand{\insertwidesrc}[3]{\begin{figure*}\centering\lstinputlisting{#1}\caption{#2}\label{fig:#3}\end{figure*}}
\newcommand{\us}{\char`_}
\newcommand{\la}{\langle}
\newcommand{\ra}{\rangle}
\newcommand{\before}{\prec}
\newcommand{\beforequal}{\preceq}
\newcommand{\concurrent}{\approx}
\newcommand{\realtime}[2]{\Gamma_{#1}(T_#2)}
\newcommand{\vclock}[2]{\la #1, #2 \ra}
\newcommand{\vclockthree}[3]{\la #1, #2, #3 \ra}
\newcommand{\ct}[1]{~\cite{#1}}
\newcommand{\titanratiotao}{10.9}
\newcommand{\titanratiow}{1.5}
\newcommand{\graphlabratio}{4}
\newcommand{\bcratio}{8}
\newcommand{\tightpara}[1]{\noindent\textbf{#1}:}
\newcommand{\sysname}{Weaver}
\newcommand{\sysnamelower}{weaver}
\newcommand{\store}{database}
\newcommand{\Store}{Database}
\newcommand{\datastore}{database}
\hyphenation{time-stamp time-stamping time-stamps time-stamper data-base
correct-ness}

\expandafter\def\expandafter\normalsize\expandafter{%
    \normalsize
    \setlength\abovedisplayskip{-400pt}
    \setlength\belowdisplayskip{-400pt}
    \setlength\abovedisplayshortskip{-400pt}
    \setlength\belowdisplayshortskip{-400pt}
}
\setlength{\floatsep}{3pt plus 1pt minus 1pt}
\setlength{\textfloatsep}{3pt plus 1pt minus 1pt}
\setlength{\intextsep}{3pt plus 1pt minus 1pt}

\title{\sysname{}: A High-Performance, Transactional Graph \Store{} Based on
Refinable Timestamps}

\numberofauthors{4}

\author{
%
%
\alignauthor
Ayush Dubey\\
       \affaddr{Cornell University}\\
\alignauthor
Greg D. Hill\\
       \affaddr{Stanford University}\\
\alignauthor
Robert Escriva\\
       \affaddr{Cornell University}\\
\and  
\alignauthor
Emin G\"un Sirer\\
       \affaddr{Cornell University}\\
}

\maketitle

\begin{abstract}
Graph \store{}s have become a common infrastructure component.
Yet existing systems either operate on offline snapshots, provide weak
consistency guarantees, or use expensive concurrency control
techniques that limit performance.

In this paper, we introduce a new distributed graph \store{}, called \sysname{},
which enables efficient, transactional graph analyses as well as strictly
serializable ACID transactions on dynamic graphs.  The key insight that
allows \sysname{} to combine strict serializability with horizontal scalability
and high performance is a novel request ordering mechanism called refinable
timestamps.  This technique couples coarse-grained vector timestamps with a
fine-grained timeline oracle to pay the overhead of strong consistency only when
needed.  Experiments show that \sysname{} enables a Bitcoin blockchain explorer
that is $\bcratio\times$ faster than Blockchain.info, and achieves
$\titanratiotao\times$ higher throughput than the Titan graph database on social
network workloads and $\graphlabratio\times$ lower latency than GraphLab on
offline graph traversal workloads.  \end{abstract}

\vspace{-\baselineskip}
\section{Introduction}\label{sec:intro}
Graph-structured data arises naturally in a wide range of fields that span
science, engineering, and business.  Social networks, the world wide web,
biological interaction networks, knowledge graphs, cryptocurrency transactions,
and many classes of business analytics are naturally modeled as a set of
vertices and edges that comprise a graph.  Consequently, there is a growing need
for systems which can store and process large-scale graph-structured data.

Correctness and consistency in the presence of changing data is a key challenge
for graph \store{}s.  For example, imagine a graph \store{} used to implement a
network controller that stores the network topology shown in
\figref{reach_prob}.  When the network is undergoing churn, it is possible for a
path discovery query to return a path through the network that did not exist at
any instant in time.  For instance, if the link $(n_3, n_5)$ fails, and
subsequently the link $(n_5, n_7)$ goes online, a path query starting from host
$n_1$ to host $n_7$ may erroneously conclude that $n_7$ is reachable from $n_1$,
even though no such path ever existed.

Providing strongly consistent queries is particularly challenging for graph
\store{}s because of the unique characteristics of typical graph queries.
Queries such as traversals often read a large portion of the graph, and
consequently take a long time to execute.  For instance, the average degree of
separation in the Facebook social network is 3.5\ct{fb_degree}, which implies that
a breadth-first traversal that starts at a random vertex and traverses 4 hops
will likely read all 1.59 billion users.  On the other hand, typical key-value
and relational queries are much smaller; the \texttt{NewOrder} transaction in
the TPC-C benchmark\ct{tpcc}, which comprises 45\% of the frequency
distribution, consists of 26 reads and writes on average\ct{warp}.  Techniques
such as optimistic concurrency control or distributed two-phase locking result
in poor throughput when concurrent queries try to read large subsets of the
graph.

Due to the unique nature of typical graph-structured data and queries, existing
databases have offered limited support.  State-of-the-art transactional graph
\store{}s such as Neo4j\ct{neo4j} and Titan\ct{titan} employ heavyweight
coordination techniques for transactions.  Weakly consistent online graph
\store{}s\ct{tao,kineo} forgo strong semantics for performance,
which limits their scope to applications with loose consistency needs and
requires complicated client logic.  Offline graph processing
systems\ct{powergraph,giraph,pregel,gps} do not permit updates to the graph
while processing queries.  Lightweight techniques for modifying and querying a
distributed graph with strong consistency guarantees have proved elusive thus
far.

\sysname{}\footnote{\href{http://weaver.systems}{http://weaver.systems},
\href{https://github.com/dubey/weaver}{https://github.com/dubey/weaver}} is a
new online, distributed, and transactional graph \store{} that supports
efficient graph analyses.  The key insight that enables \sysname{} to scalably
execute graph transactions in a strictly serializable order is a novel technique
called \emph{refinable timestamps}.  This technique uses a highly scalable and
lightweight timestamping mechanism for ordering the majority of operations and
relies on a fine-grained timeline oracle for ordering the remaining,
potentially-conflicting reads and writes.  This unique two-step ordering
technique with proactive timestamping and a reactive timeline oracle has three
advantages.

\insertfig{0.56}{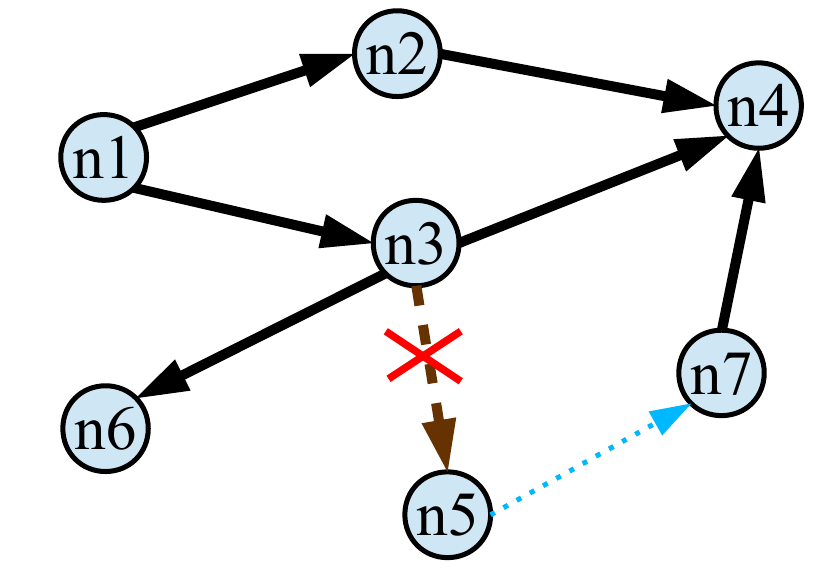}{A graph undergoing an update which
creates $(n_5, n_7)$ and deletes $(n_3, n_5)$ concurrently with a traversal
starting at $n_1$.  In absence of transactions, the query can return path $(n_1,
n_3, n_5, n_7)$ which never existed.}{reach_prob}

First, refinable timestamps enable \sysname{} to distribute the graph across
multiple shards and still execute transactions in a scalable fashion.  There are
some applications and workloads for which sharding is
unnecessary\ct{laptop_graph}.  However many applications support a large number
of concurrent clients and operate on graphs of such large scale, consisting of
billions of vertices and edges\ct{fb_anatomy,twitter_rv,bitcoin}, that a
single-machine architecture is infeasible.  For such high-value
applications\ct{tao,robobrain} it is critical to distribute the graph data in
order to balance the workload and to enable highly-parallel in-memory query
processing by minimizing disk accesses.

Second, refinable timestamps reduce the amount of coordination required for
execution of graph analysis queries.  Concurrent transactions that do not
overlap in their data sets can execute independently without blocking each
other.  Refinable timestamps order only those transactions that overlap in their
read-write sets, using a combination of vector clock ordering and the timeline
oracle.

Third, refinable timestamps enable \sysname{} to store a multi-version graph by
marking vertices and edges with the timestamps of the write operations. A
multi-version graph lets long-running graph analysis queries operate on a
consistent version of the graph without blocking concurrent writes. It also
allows historical queries which run on past, consistent versions of the graph.

Overall, this paper makes the following contributions:
\begin{itemize}
    \setlength{\itemsep}{3pt}
    \item It describes the design of an online, distributed, fault-tolerant, and
        strongly consistent graph \store{} that achieves high performance and
        enables ACID transactions and consistent graph queries.

    \item It details a novel, lightweight ordering mechanism called refinable
        timestamps that enables the system to trade off proactive
        synchronization overheads with reactive discovery of ordering
        information.

    \item It shows that these techniques are practical through a full
        implementation and an evaluation that shows that \sysname{} scales well
        to handle graphs with over a billion edges and significantly outperforms
        state-of-the-art systems such as Titan\ct{titan} and
        GraphLab\ct{powergraph} and applications such as
        Blockchain.info\ct{bcinfo}.
\end{itemize}

\vspace{-0.5\baselineskip}
\section{Approach}\label{sec:approach}
\sysname{} combines the strong semantics of ACID transactions with
high-performance, transactional graph analyses.  In this section, we describe
the data and query model of the system as well as sample applications.

\subsection{Data Model}\label{sec:data_model}
\sysname{} provides the abstraction of a property graph, i.e. a directed graph
consisting of a set of vertices with directed edges between them.  Vertices and
edges may be labeled with named properties defined by the application.  For
example, an edge $(u,v)$ may have both ``weight=3.0'' and ``color=red''
properties, while another edge $(v,w)$ may have just the ``color=blue''
property.  This enables applications to attach data to vertices and edges.

\subsection{Transactions for Graph Updates}
\sysname{} provides transactions over the directed graph abstraction.  These
transactions comprise reads and writes on vertices and edges, as well as their
associated attributes.  The operations are encapsulated in a
\texttt{\sysnamelower{}\us tx} block and may use methods such as \texttt{get\us
vertex} and \texttt{get\us edge} to read the graph, \texttt{create/delete\us
vertex} and \texttt{create/delete\us edge} to modify the graph structure, and
\texttt{assign/delete\us properties} to assign or remove attribute data on
vertices and edges.  \srcref{socnet_write} shows the code for an update to a
social network that posts content and manages the access control for that
content in the same atomic transaction.

\insertsrc{src/socnet_write.py}{A \sysname{} transaction which posts a photo in
a social network and makes it visible to a subset of the user's
friends.}{socnet_write}

\subsection{Node Programs for Graph Analyses}
\sysname{} also provides specialized, efficient support for a class of read-only
graph queries called node programs.  Similar to stored procedures in
databases\ct{dbms_stored_proc}, node programs traverse the graph in an
application-specific fashion, reading the vertices, edges, and associated
attributes via the \texttt{node} argument.  For example, \srcref{reach_prog}
describes a node program that executes BFS using only edges annotated with a
specified \texttt{edge\_property}.  Such queries operate atomically and in
isolation on a logically consistent snapshot of the graph.  \sysname{} queries
wishing to modify the graph must collate the changes they wish to make in a node
program and submit them as a transaction.

\sysname{}'s node programs employ a mechanism similar to the commonly used
scatter-gather approach\ct{pregel,giraph,powergraph} to propagate queries to
other vertices.  In this approach, each vertex-level computation is passed query
parameters (\texttt{prog\us params} in \srcref{reach_prog}) from the previous
hop vertex, similar to the gather phase.  Once a node program completes
execution on a given vertex, it returns a list of vertex handles to traverse
next, analogous to the scatter phase.  A node program may visit a vertex any
number of times; \sysname{} enables applications to direct all aspects of node
program propagation.  This approach is sufficiently expressive to capture common
graph analyses such as graph exploration\ct{bcinfo}, search
algorithms\ct{fb_graphsearch}, and path discovery\ct{robobrain}.

Many node programs are stateful.  For instance, a traversal query may store a
bit per vertex visited, while a shortest path query may require state to save
the distance from the source vertex.  This per-query state is represented in
\sysname{}'s node programs by \texttt{node.prog\us state}.  Each active node
program has its own state object that persists within the \texttt{node} object
until the node program runs to completion throughout the graph.  As a node
program traverses the graph, the application can create \texttt{prog\us state}
at other vertices and propagate it between vertices using the \texttt{prog\us
params}.  This design enables applications that implement a wide array of graph
algorithms.  Node program state is garbage collected after the query terminates
on all servers (\secref{gc}).

Since node programs are typically long-running, it is a challenge to ensure that
these queries operate on a consistent snapshot of the graph.  \sysname{}
addresses this problem by storing a multi-version graph with associated
timestamps.  This enables transactional graph updates to proceed without
blocking on node program reads.

\insertsrc{src/reach_program.py}{A node program in \sysname{} which executes a BFS
query on the graph.}{reach_prog}

\vspace{-\baselineskip}
\section{Refinable Timestamps}\label{sec:rt_intro}
The key challenge in any transactional system is to ensure that distributed
operations taking place on different machines follow a coherent timeline.
\sysname{} addresses this challenge with refinable timestamps, a lightweight
mechanism for achieving a rough order when sufficient and fine-grained order
when necessary.

\subsection{Overview}
At a high level, refinable timestamps factor the task of achieving a strictly
serializable order of transaction execution into two stages.  The first stage,
which assigns a timestamp to each transaction, is cheap but imprecise.  Any
server in the system that receives the transaction from a client can assign the
timestamp, without coordinating with other servers.  There is no distributed
coordination, resulting in high scalability.  However, timestamps assigned in
this manner are imprecise and do not give a total order between transactions.

The second stage resolves conflicts that may arise during execution of
transactions with imprecise timestamps.  This stage is more expensive and less
scalable but leads to a precise ordering of transactions.  The system resorts to
the second stage only for a small subset of transactions, i.e. those that are
concurrent and overlap in their read-write sets.

The key benefit of using refinable timestamps, compared to traditional
distributed locking techniques, is reduced coordination.  The proactive stage is
lightweight and scalable, and imposes very little overhead on transaction
processing.  The system pays the cost of establishing a total order only when
conflicts arise between timestamped operations.  Thus, refinable timestamps
avoid coordinating transactions that do not conflict.

This benefit is even more critical for a graph \store{} because of the
characteristics of graph analysis queries: long execution time and large read
set.  For example, a breadth-first search traversal can explore an expansive
connected component starting from a single vertex.  Refinable timestamps execute
such large-scale reads without blocking concurrent, conflicting transactions.

\insertfig{0.98}{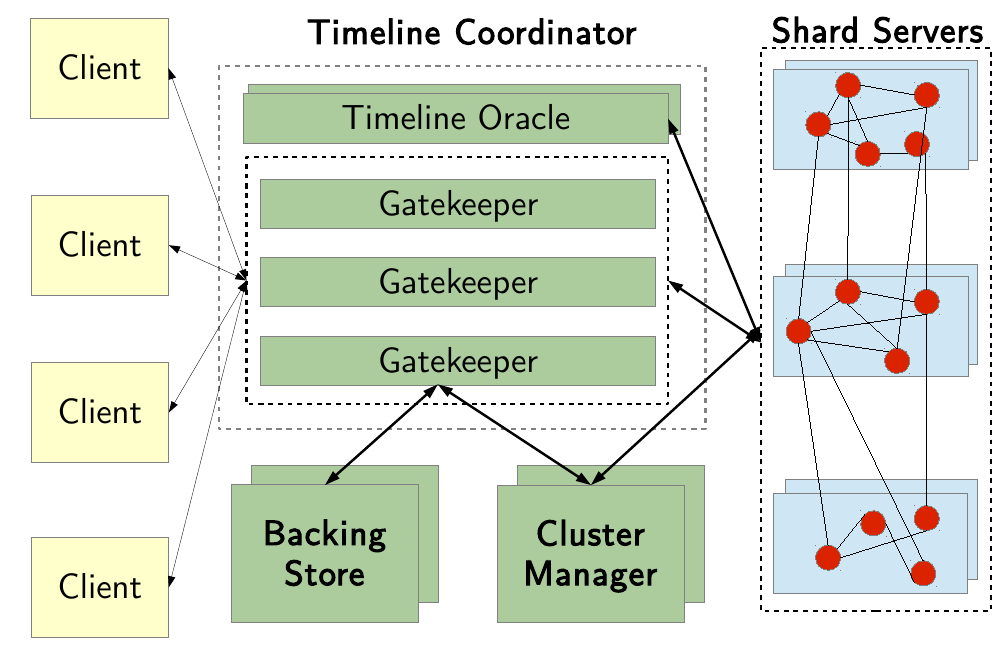}{\sysname{} system architecture.}{arch}

\subsection{System Architecture}\label{sec:arch}
\sysname{} implements refinable timestamps using a timeline coordinator, a set
of shard servers and a backing store.  \figref{arch} depicts the \sysname{}
system architecture.

\tightpara{Shard Servers}
\sysname{} distributes the graph by partitioning it into smaller pieces, each of
which is stored in memory on a shard server.  This sharding enables both memory
storage capacity and query throughput to scale as servers are added to the
system.  Each graph partition consists of a set of vertices, all outgoing edges
rooted at those vertices, and associate attributes.  The shard servers are
responsible for executing both node programs and transactions on the in-memory
graph data.

\tightpara{Backing Store}
The backing store is a key-value store that supports ACID transactions and
serves two purposes.  First, it stores the graph data in a durable and
fault-tolerant manner.  When a shard server fails, the graph data that belongs
to the shard is recovered from the backing store.  Second, the backing store
directs transactions on a vertex to the shard server responsible for that vertex
by storing a mapping from vertices to associated shard servers.  Our
implementation uses HyperDex Warp\ct{warp} as the backing store.

\tightpara{Timeline Coordinator}
The critical component behind \sysname{}'s strict serializability guarantees is
the timeline coordinator.  This coordinator consists of a user-configured number
of \emph{gatekeeper} servers for coarse timestamp-based ordering and a
\emph{timeline oracle} for refining these timestamps when necessary
(\secref{coarse_grain}, \secref{fine_grain}).  In addition to assigning
timestamps to transactions, the gatekeepers also commit transactional updates to
the backing store (\secref{impl}).

\tightpara{Cluster Manager}
\sysname{} also deploys a cluster manager process for failure detection and
system reconfiguration.  The cluster manager keeps track of all shard servers
and gatekeepers that are currently part of the \sysname{} deployment.  When a
new gatekeeper or shard server boots up, it registers its presence with the
cluster manager and then regularly sends heartbeat messages.  If the cluster
manager detects that a server has failed, it reconfigures the cluster according
to \sysname{}'s fault tolerance scheme (\secref{ft}).

\vspace{-\baselineskip}
\subsection{Proactive Ordering by Gatekeepers}\label{sec:coarse_grain}
The core function of gatekeepers is to assign to every transaction a timestamp
that can scalably achieve a partial order.  To accomplish this, \sysname{}
directs each transaction through any one server in a bank of gatekeepers, each
of which maintains a vector clock\ct{fidge_timestamps}.  A vector clock consists
of an array of counter values, one per gatekeeper, where each gatekeeper
maintains a local counter as well as the maximum counter value it has seen from
the other gatekeepers.  Gatekeepers increment their local clock on receipt of a
client request, attach the vector clock to every such transaction, and forward
it to the shards involved in the transaction.

Gatekeepers ensure that the majority of transaction timestamps are directly
comparable by exchanging vector clocks with each other every $\tau$
milliseconds.  This proactive communication between gatekeepers establishes a
\emph{happens-before} partial order between refinable timestamps.
\figref{gk_time} shows how these vector clocks can order transactions with the
help of these happens-before relationships.  In this example, since $T_1$ and
$T_2$ are separated by an announce message from gatekeeper 0, their vector
timestamps are sufficient to determine that $T_1 \vclockthree{1}{1}{0} \before
T_2 \vclockthree{3}{4}{2}$ (X $\before$ Y denotes X happens before Y, while X
$\beforequal$ Y denotes either X $\before$ Y or X $=$ Y).

\insertfig{0.8}{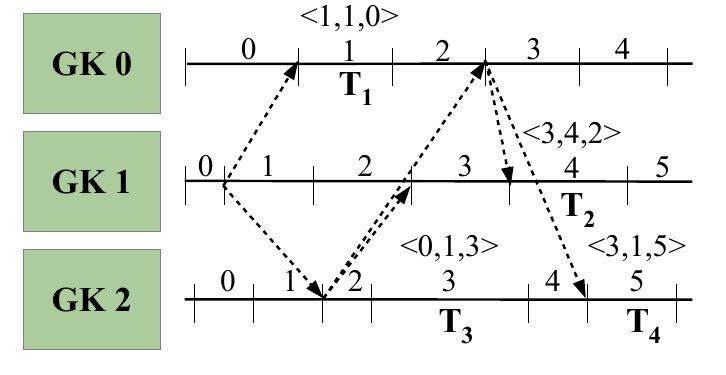}{Refinable timestamps using three
    gatekeepers.  Each gatekeeper increments its own counter for a transaction
    and periodically announces its counter to other gatekeepers (shown
    by dashed arrows).  Vector timestamps are assigned locally based on
    announcements that a gatekeeper has collected from peers.  $T_1
    \vclockthree{1}{1}{0} \before T_2 \vclockthree{3}{4}{2}$ and $T_3
    \vclockthree{0}{1}{3} \before T_4 \vclockthree{3}{1}{5}$.  $T_2$ and $T_4$
    are concurrent and require fine-grain ordering only if they conflict. There
    is no need for lockstep synchrony between gatekeepers.}{gk_time}

Unfortunately, vector clocks are not sufficient to establish a total order.  For
instance, in \figref{gk_time}, transactions $T_2$ (with timestamp
$\vclockthree{3}{4}{2}$) and $T_4$ (with timestamp $\vclockthree{3}{1}{5}$)
cannot be ordered with respect to each other and need a more refined ordering if
they overlap in their read-write sets (we denote this as $T_2 \concurrent T_4$).
Since transactions that enter the system simultaneously through multiple
gatekeepers may receive concurrent vector clocks, \sysname{} uses an auxiliary
service called a timeline oracle to put them into a serializable timeline.

\vspace{-\baselineskip}
\subsection{Reactive Ordering by Timeline Oracle}\label{sec:fine_grain}
A timeline oracle is an event ordering service that keeps track of the
happens-before relationships between transactions\ct{kronos}.  The timeline
oracle maintains a dependency graph between outstanding transactions, completely
independent of the graph stored in \sysname{}.  Each vertex in the dependency
graph represents an ongoing transaction, identified by its vector timestamp, and
every directed edge represents a happens-before relationship.  The timeline
oracle ensures that transactions can be reconciled with a coherent timeline by
guaranteeing that the graph remains acyclic. 

The timeline oracle carries out this task with a simple API centered around
events, where an event corresponds to a transaction in \sysname{}. Specifically,
it provides primitives to create a new event, to atomically assign a
happens-before relationship between sets of events, and to query the order
between two or more events. 

\sysname{}'s implementation of the timeline oracle comprises such an
event-oriented API backed by an event dependency graph that keeps track of
transactions at a fine grain\ct{kronos}.  The service is essentially a state
machine that is chain replicated\ct{chain_rep} for fault tolerance.  Updates to
the event dependency graph, caused by new events or new dependencies, occur at
the head of the chain, while queries can execute on any copy of the graph.  This
results in a high-performance implementation that scales up to $\sim$6M queries
per second on a 12 8-core server chain.

\sysname{} uses this high-performance timeline oracle to establish an
order between concurrent transactions which may overlap in their read or write
sets.  Strictly speaking, such transactions must have at least one vertex or
edge in common.  Since discovering fine-grained overlaps between transaction
operations can be costly, our implementation conservatively orders any pair of
concurrent transactions that have a shard server in common.  When two such
transactions are committing simultaneously, the server(s) committing the
transactions send an ordering request to the timeline oracle.  The oracle either
returns an order if it already exists, or establishes an order between the
transactions.  To maintain a directed acyclic graph corresponding to the
happens-before relationships, it ensures that all subsequent operations follow
this order.

Establishing a fine-grained order on demand has the significant advantage that
\sysname{} will not order transactions that cannot affect each other, thereby
avoiding the overhead of the centralized oracle for these transactions
(\secref{impl_nodeprog}, \secref{impl_tx}).  Such transactions will commit
without coordination. Their operations may interleave, i.e.  appear non-atomic
to an omniscient observer, but this interleaving is benign because, by
definition, no clients can observe this interleaving.  The only transactions
that need to be ordered are those whose interleaving may lead to an observable
non-atomic or non-serializable outcome.

\vspace{-\baselineskip}
\subsection{Discussion}\label{sec:timeline_summary}
\sysname{}'s implementation of refinable timestamps combines vector clocks with
a timeline oracle.  Alternatively, the gatekeepers in \sysname{} could assign a
loosely synchronized real timestamp to each transaction, similar to
TrueTime\ct{spanner}.  Both techniques ensure a partial order.  However TrueTime
makes assumptions about network synchronicity and communication delay, which are
not always practical, even within the confines of a datacenter.  Synchronicity
assumptions interfere with debugging, and maybe violated by network delays under
heavy load and systems running in virtualized environments.  Further, a TrueTime
system synchronized with average error bound $\bar\varepsilon$ will necessarily
incur a mean latency of $2\bar\varepsilon$. While TrueTime makes sense for the
wide area environment for which it was developed, \sysname{} uses vector clocks
for its first stage.


Irrespective of implementation, refinable timestamps represent a hybrid approach
to timeline ordering that offers an interesting tradeoff between proactive costs
due to periodic synchronization messages between gatekeepers, and the reactive
costs incurred at the timeline oracle.  At one extreme, one could use the
timeline oracle for maintaining the global timeline for \emph{all} requests, but
then the throughput of the system would be bottlenecked by the throughput of the
oracle.  At the other extreme, one could use only gatekeepers and synchronize at
such high frequency so as to provide no opportunity for concurrent timestamps to
arise.  But this approach would also incur too high an overhead, especially
under high workloads.  \sysname{}'s key contribution is to reduce the load on a
totally ordering timeline oracle by layering on a timestamping service that
manages the bulk of the ordering, and leaves only a small number of overlapping
transactions to be ordered by the oracle.  This tradeoff ensures that the
scalability limits of a centralized timeline service\ct{kronos} are extended by
adding gatekeeper servers.  \sysname{}'s design provides a parameter
$\tau$---the clock synchronization period---that manages this tradeoff.  

The clock synchronization period can be adjusted dynamically based on the system
workload.  Initially, when the system is quiescent, the gatekeepers do not need
to synchronize their clocks.  As the rate of transactions processed by the
different gatekeepers increases, the gatekeepers synchronize clocks more
frequently to reduce the burden on the timeline oracle.  Beyond a point, the
overhead of synchronization itself reduces the throughput of the timestamping
process.  We empirically analyze how the system can discover the sweet spot for
$\tau$ in \secref{eval}.

\vspace{-0.5\baselineskip}
\section{Implementation and \\ Correctness}\label{sec:impl}
\sysname{} uses refinable timestamps for ordering transactions.  However,
because node programs potentially have a very large read set and long execution
time, \sysname{} processes node programs differently from read-write
transactions.

\subsection{Node Programs}\label{sec:impl_nodeprog}
\sysname{} includes a specialized, high-throughput implementation of refinable
timestamps for node program execution.  A gatekeeper  assigns a timestamp
$T_{prog}$ and forwards the node program to the appropriate shards.  The shards
execute the node program on a version of the in-memory graph consistent with
$T_{prog}$ by comparing $T_{prog}$ to the timestamps of the vertices and edges
in the multi-version graph and only reading the portions of the graph that exist
at $T_{prog}$.  In case timestamps are concurrent, the shard requests for an
order from the timeline oracle.

When the timeline oracle receives an ordering request for a node program and a
committed write from a shard, it returns the pre-established order between these
transactions to the shard, if one exists. In cases where a pre-established order
does not exist, because gatekeepers do not precisely order transactions, the
oracle will prefer arrival order.  This order is then established as a
commitment for all time; the timeline oracle will record the happens-before
relationship and ensure that all subsequent queries from all shard servers
receive responses that respect this commitment.

Because arrival order may differ on different shard servers, care must be taken
to ensure atomicity and isolation.  For example, in a na\"ive implementation, a
node program $P$ may arrive after a transaction $T$ on shard 2, but before $T$
on shard 1.  To ensure consistent ordering, \sysname{} delays execution of a
node program at a shard until after execution of all preceding and concurrent
transactions.

In addition to providing consistent ordering for transactions, the timeline
oracle ensures that transitive ordering is maintained. For
instance, if $T_1 \before T_2$ and $T_2 \before T_3$ is pre-established, then an
order query between $T_1$ and $T_3$ will return $T_1 \before T_3$.
Furthermore, because transactions are identified by their unique vector clocks,
the timeline oracle can infer and maintain implicit dependencies captured by
the vector clocks. For example, if the oracle first orders $\vclock{0}{1}
\before \vclock{1}{0}$ and subsequently a shard requests the order between
$\vclock{0}{1}$ and $\vclock{2}{0}$, the oracle will return $\vclock{0}{1}
\before \vclock{2}{0}$ because $\vclock{0}{1} \before \vclock{1}{0} \before
\vclock{2}{0}$ due to transitivity.

\subsection{Transactions}\label{sec:impl_tx}
Transactions, which contain both reads and writes, result in updates to both the
in-memory graph at the shard servers and the fault-tolerant graph stored in the
backing store.  \sysname{} first executes the transaction on the backing store,
thereby leveraging its transactional guarantees to check transaction validity.
For example, if a transaction attempts to delete an already deleted vertex, it
aborts while executing on the backing store.  After the transaction commits
successfully on the backing store, it is forwarded to the shard servers which
update the in-memory graph without coordination.  

To execute a transaction on the backing store, gatekeepers act as
intermediaries.  Clients buffer writes and submit them as a batch to the
gatekeeper at the end of a transaction, and the gatekeeper, in turn, performs
the writes on the backing store.  The backing store commits the transaction if
none of the data read during the transaction was modified by a concurrent
transaction.  HyperDex Warp, the backing store used in Weaver, employs the
highly scalable acyclic transactions protocol\ct{warp} to order multi-key
transactions.  This protocol a form optimistic concurrency control that enables
scalable execution of large volumes of transactions from gatekeepers.

Gatekeepers, in addition to executing transactions on the backing store, also
assign a refinable timestamp to each transaction.  Timestamps are assigned in a
manner that respects the order of transaction execution on the backing store.
For example, if there are two concurrent transactions $T_1$ and $T_2$ at
gatekeepers $GK_1$ and $GK_2$ respectively, both of which modify the same vertex
in the graph, \sysname{} guarantees that if $T_1$ commits before $T_2$ on the
backing store, then $T_1 \before T_2$.  To this end, \sysname{} stores the
timestamp of the last update for each vertex in the backing store.  In our
example, if $T_1$ commits before $T_2$ on the backing store, then the last
update timestamp at the graph vertex will be $T_1$ when $GK_2$ attempts to
commit $T_2$.  Before committing $T_2$, $GK_2$ will check that $T_1 \before
T_2$.  If it so happens that the timestamp assigned by $GK_2$ is smaller, i.e.
$T_2 \before T_1$, then $GK_2$ will abort and the client will retry the
transaction.  Upon retrying, $GK_2$ will assign a higher timestamp to the
transaction.

\insertfig{0.7}{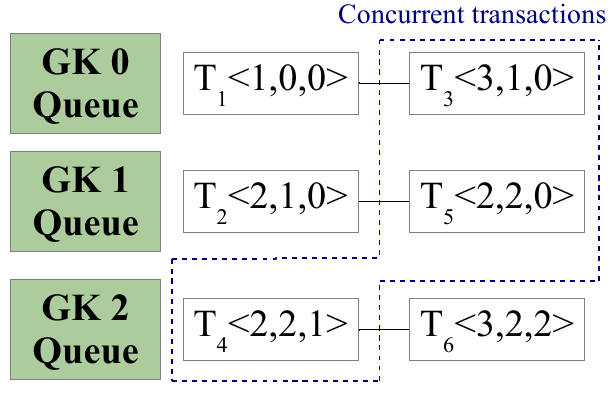}{Each shard server maintains a queue of
transactions per gatekeeper and executes the transaction with the lowest
timestamp. When a group of transactions are concurrent (e.g.  $T_3, T_4,$ and
$T_5$), the shard server consults the timeline oracle to order them.}{queues}

While gatekeepers assign refinable timestamps to transactions and thereby
establish order, shard servers obey this order.  To do so, each shard server has
a priority queue of incoming transactions for each gatekeeper, prioritized by
their timestamps (\figref{queues}).  Shard servers enqueue transactions from
gatekeeper $i$ on its $i$-th \emph{gatekeeper queue}.  When each gatekeeper
queue is non-empty, an event loop at the shard server pulls the first
transaction $T_i$ off each queue $i$ and executes the earliest transaction out
of $(T_1, T_2, \ldots, T_n)$.  In case a set of transactions appear concurrent,
such as $(T_3, T_4, T_5)$ in \figref{queues}, the shard servers will submit the
set to the timeline oracle in order to discover and, if necessary, assign an
order.


\sysname{}'s implementation of refinable timestamps at shard servers has
correctness and performance subtleties.  First, in order to ensure that
transactions are not lost or reordered in transit,
\sysname{} maintains FIFO channels between each gatekeeper and shard pair using
sequence numbers.  Second, to ensure the system makes progress in periods of
light workload, gatekeepers periodically send NOP transactions to shards.  NOP
transactions guarantee that there is always a transaction at the head of each
gatekeeper queue.  This provides an upper-bound on the delay in node program
execution, set by default to $10 \mu s$ in our current implementation.
Finally, since ordering decisions made by the timeline oracle are irreversible
and monotonic, shard servers can cache these decisions in order to reduce the
number of ordering requests.


Shard servers also maintain the in-memory, multi-version distributed graph by
marking each written object with the refinable timestamp of the transaction.
For example, an operation that deletes an edge actually marks the edge as
deleted and stores the refinable timestamp of the deletion in the edge object.

\vspace{-0.5\baselineskip}
\subsection{Fault Tolerance}\label{sec:ft}
\sysname{} minimizes data that is persistently stored by only storing the the
graph data in the backing store.  In response to a gatekeeper or shard failure,
the cluster manages spawns a new process on one of the live machines.  The new
process restores corresponding graph data from the backing store.  However,
restoring the graph from the backing store is not sufficient to ensure strict
serializability of transactions since timestamps and queues are not stored at
the backing store.

\sysname{} implements additional techniques to ensure strict serializability.
A transaction that has committed on the backing store
before failure requires no extra handling: the backup server will read the
latest copy of the data from the backing store.  Transactions that have not
executed on the backing store, as well as all node programs, are reexecuted by
\sysname{} with a fresh timestamp after recovery, when resubmitted by clients.
Since partially executed operations for these transactions were not
persistent, it is safe to start execution from scratch.  This simple strategy
avoids the overhead of execution-time replication of ordering metadata such as
the gatekeeper queues at shards and pays the cost of reexecution on rare
server failures.

Finally, in order to maintain monotonicity of timestamps on gatekeeper failures,
a backup gatekeeper restarts the vector clock for the failed gatekeeper.  To
order the new timestamps with respect to timestamps issued before the failure,
the vector clocks in \sysname{} include an extra \emph{epoch} field which the
cluster manager increments on failure detection.  The cluster manager imposes a
barrier between epochs to guarantee that all servers move to the new epoch in
unison.

The cluster manager and the timeline oracle are fault-tolerant Paxos\ct{paxos}
replicated state machines\ct{rsm}.

\subsection{Proof of Correctness}\label{sec:proof}
In this section, we prove that \sysname{}'s implementation of refinable
timestamps yields a strictly serializable execution order of transactions and
node programs.  We structure the proof in two parts---the first part shows that
the execution order of transactions is serializable, and the second part shows
that the execution order respects wall-clock ordering.  We assume that the
timeline oracle correctly maintains a DAG of events that ensures that no cycles
can arise in the event dependency graph\ct{kronos}.

\newtheorem{serial}{Strict Serializability}

\vspace{-0.5\baselineskip}
\begin{serial}
    Let transactions $T_1, T_2, \ldots, T_n$ have timestamps $t_1, t_2, \ldots,
    t_n$.  Then the execution order of $T_1, T_2, \ldots, T_n$ in \sysname{} is
    equivalent to a serializable execution order.
\end{serial}
\vspace{-\baselineskip}
\begin{proof}
    We prove the claim by induction on $n$, the number of transactions in the
    execution.

    \textbf{Basis}: The case of $n=1$, the execution of a single transaction
    $T_1$ is vacuously serializable.

    \textbf{Induction}: Assume all executions with $n$ transactions are
    serializable in \sysname{}.  Consider an execution of $n+1$ transactions.
    Remove any one transaction from this execution, say $T_i, 1 \le i \le n+1$,
    resulting in a set of $n$ transactions.  The execution of these transactions
    has an equivalent serializable order because of the induction hypothesis.
    We will prove that the addition of $T_i$ to the execution also yields a
    serializable order by considering the ordering of $T_i$ with an arbitrary
    transaction $T_j, 1 \le j \le n+1, i \neq j$ in three cases.

    First, if both $T_i$ and $T_j$ are node programs, then their relative
    ordering does not matter as they do not modify the graph data.

    Second, let $T_i$ be a node program and $T_j$ be a read-write transaction.
    If $T_i \before T_j$, either due to vector clock ordering or due to the
    timeline oracle, then the node program $T_i$ cannot read any of $T_j$'s
    updates.  This is because when $T_i$ executes at a vertex $v$, \sysname{}
    first iterates through the multi-version graph data (i.e. vertex properties,
    out-edges, and edge properties) associated with $v$, and filters out updates
    that happen after $t_i$ (\secref{impl_nodeprog}).  If $T_j \before T_i$,
    then \sysname{} ensures that $T_i$ reads all updates, across all shards, due
    to $T_j$.  This is because node program execution is delayed at a shard
    until the timestamp of the node program is lower than all enqueued
    read-write transactions (\secref{impl_nodeprog}).

    Third, we consider the case when both $T_i$ and $T_j$ are read-write
    transactions.  Let $\realtime{x}{k}$ denote the real time of execution of
    transaction $T_k$ at shard $S_x$.  If $t_i < t_j$ due to vector clock
    ordering, then $\realtime{x}{i} < \realtime{x}{j} \; \forall x$.
    (\secref{impl_tx}).  Similarly if $t_j < t_i$ then $\realtime{x}{j} <
    \realtime{x}{i} \; \forall x$.  For the case when $t_i \concurrent t_j$,
    assume if possible that $T_i$ and $T_j$ are not consistently ordered across
    all shards, i.e.  $\realtime{a}{i} < \realtime{a}{j}$ and $\realtime{b}{j} <
    \realtime{b}{i}$.  When $T_j$ executes at $S_b$, let $T_i'$ be the
    transaction that is in the gatekeeper queue corresponding to $T_i$.  $T_i'$
    may either be the same as $T_i$, or $T_i' \before T_i$ due to sequence
    number ordering (\secref{impl_tx}).  Since $\realtime{b}{j} <
    \realtime{b}{i'}$, we must have $T_j \before T_i'$. But since $t_i
    \concurrent t_j$, it must also be the case that $t_i' \concurrent t_j$, and
    thus the decision $T_j \before T_i'$ was established at the timeline oracle.
    Thus we have:
    \vspace{-2pt}
    \begin{equation} \label{eq:1}
        T_j \before T_i' \beforequal T_i
    \end{equation}
    \vspace{-2pt}
    Now when $T_i$ executes at $S_a$, let $T_j'$ be the transaction in the
    gatekeeper queue corresponding to $T_j$.  By an argument identical to the
    previous reasoning, we get:
    \vspace{-2pt}
    \begin{equation} \label{eq:2}
        T_i \before T_j' \beforequal T_j
    \end{equation}
    \vspace{-2pt}
    \eqnref{1} and \eqnref{2} yield a cycle in the dependency graph, which is
    not permitted by the timeline oracle.

    Since the execution of $T_i$ is isolated with respect to the execution of an
    arbitrary transaction $T_j \; \forall j, 1 \le j \le n+1$, we can insert $T_i$
    in the serial execution order of $T_1, \ldots, T_{i-1},$ $T_{i+1},
    \ldots, T_{n+1}$ and obtain another serializable execution order comprising
    all $n+1$ transactions.
\end{proof}

\vspace{-0.5\baselineskip}
\begin{serial}
    Let transactions $T_1$ and $T_2$ have timestamps $t_1$ and $t_2$
    respectively.  If the invocation of transaction $T_2$ occurs after the
    response for transaction $T_1$ is returned to the client, then \sysname{}
    orders $T_1 \before T_2$.
\end{serial}
\begin{proof}
    When both $T_1$ and $T_2$ are read-write transactions, then the natural
    execution order of the transactions on the transactional backing store
    ensures that $T_1 \before T_2$.  This is because the response of $T_1$ is
    returned to the client only after the transaction executes on the backing
    store, and the subsequent invocation of $T_2$ will see the effects of $T_1$.

    Consider the case when the invocation of node program $T_2$ occurs after the
    response of transaction $T_1$.  If either $t_1 < t_2$ or $t_2 < t_1$ by
    vector clock ordering, the shards will order the node program and
    transaction in their natural timestamp order.  When $t_1 \concurrent t_1$,
    the timeline oracle will consistently order the two transactions across all
    shards.  The oracle will return a preexisting order if one exists, or order
    the node program after the transaction (\secref{impl_nodeprog}).  By always
    ordering node programs after transactions when no order exists already, the
    timeline oracle ensures that node programs never miss updates due to
    completed transactions.
\end{proof}

Combining the two theorems yields that \sysname{}'s implementation of refinable
timestamps results in a strictly serializable order of execution of transactions
and node programs.

\subsection{Garbage Collection}\label{sec:gc}
\sysname{}'s multi-version graph data model permits multiple garbage collection
policies for deleted objects.  Users may choose to not collect old state if they
wish to maintain a multi-version graph with support for historic searches, or
they may choose to clean up state older than the earliest operation still in
progress within the system.  In the latter case, gatekeepers periodically
communicate the timestamp of the oldest ongoing node program to the shards.
This allows the shards to delete all versions older than the oldest node
program.  \sysname{} uses a similar GC technique to clean up the dependency
graph of the timeline oracle.


\subsection{Graph Partitioning and Caching}\label{sec:partition_cache}
Graph queries often exhibit locality in their data access patterns: a query that
reads a vertex often also reads its neighbors.  \sysname{} leverages this by
dynamically colocating a vertex with the majority of its neighbors, using
streaming graph partitioning algorithms\ct{streaming_heur,restreaming}, to
reduce communication overhead during query processing.

\newcommand{\graphpath}[2]{(V_{#1}, \ldots, V_{#2})}
In addition to locality in access patterns, graph analyses can benefit from
caching analysis results at vertices.  For example, a path query that discovers
the path $\graphpath{1}{n}$ can cache the $\graphpath{i}{n}$ path at each vertex
$V_i$.  \sysname{} enables applications to memoize the results of node programs
at vertices and to reuse the memoized results in subsequent executions.  In
order to maintain consistency guarantees, \sysname{} enables applications to
invalidate the cached results by discovering the changes in the graph structure
since the result was cached.  Thus, in our previous example, whenever any vertex
or edge along the $\graphpath{1}{n}$ path is deleted, the application can
discard the cached value and reexecute the node program.

The details of the partitioning and caching mechanisms are orthogonal
to the implementation of refinable timestamps and are beyond the scope of this
paper.  Accordingly, we disable both mechanisms in the system evaluation.

\vspace{-0.5\baselineskip}
\section{Applications}\label{sec:apps}
\sysname{}'s property graph abstraction, together with strictly
serializable transactions, enable a wide variety of applications.  We describe
three sample applications built on \sysname{}.

\subsection{Social Network}\label{sec:socnet_app}
We implement a database backend for a social network, based on the Facebook TAO
API\ct{tao}, on \sysname{}.  Facebook uses TAO to store both their social
network as well as other graph-structured metadata such as relationship between
status updates, photos, `likes', comments, and users.  Applications attributes
vertices and edges with data that helps render the Facebook page and enable
important application-level logic such as access control.  TAO supports billions
of reads and millions of writes per second and manages petabytes of
data\ct{tao}.  We evaluate the performance of this social network backend
against a similar one implemented on Titan\ct{titan}, a popular open-source
graph database, in \secref{eval_titan}.


\subsection{CoinGraph}
Bitcoin\ct{bitcoin} is a decentralized cryptocurrency that maintains a
publicly-accessible history of transactions stored in a datastructure called the
blockchain.  For each transaction, the blockchain details the source of the
money as well as the output Bitcoin addresses.  CoinGraph is a
blockchain explorer that stores the transaction data as a directed graph in
\sysname{}.  As Bitcoin users transact, CoinGraph adds vertices and edges to
\sysname{} in real time.  CoinGraph uses \sysname{}'s node programs to execute
algorithms such as user clustering, flow analyses, and taint tracking.  The
application currently stores more than 80M vertices and 1.2B edges, resulting in
a total of $\sim 900$ GB of annotated data in \sysname{}.

\subsection{RoboBrain}
RoboBrain\ct{robobrain} stores a knowledge graph in \sysname{} that assimilates
data and machine learning models from a variety of sources, such as physical
robot interactions and the WWW, into a semantic network.  Vertices correspond to
concepts and edges represent labeled relationships between concepts.  As
RoboBrain incorporates potentially noisy data into the network, it merges this
data into existing concepts and splits existing concepts transactionally.
\sysname{} also enables RoboBrain applications to perform subgraph queries as a
node program.  This allows ML researchers to learn new concepts without worrying
about data or model inconsistencies on potentially petabytes of
data\ct{robobrain_news1}.

\subsection{Discussion}
The common theme among these applications is the need for transactions on
dynamic graph structured data.  For example, if the social network backend did
not support strictly serializable transactions, it would be possible for reads
to see an inconsistent or out-of-date view of the graph, leading to potentially
serious security flaws such as access control violations.  Indeed, Facebook
recently published a study of consistency in the TAO database\ct{fb_consistency}
which showed that in a trace of 2.7B requests over 11 days, TAO served thousands
of stale reads that violate linearizability.  Similarly, if CoinGraph were to be
built on a non-transactional database, then it would be possible for users to
see a completely incorrect view of the blockchain.  This is possible because (1)
the Bitcoin protocol accepts new transactions in blocks and partially executed
updates can lead to an inconsistent blockchain, and (2) in the event of a
blockchain fork, a database that reads a slightly stale snapshot may return
incorrect transactions from the wrong branch of the blockchain fork, leading to
financial losses.

While it may be possible to build specialized systems for some of these
applications that relax the strict serializability guarantees, we believe that
providing transactional semantics to developers greatly simplifies the design of
such applications.  Moreover, since semantic bugs are the leading cause of
software bugs\ct{bugs_survey}, a well-understood API will reduce the number of
such bugs.  Finally, \sysname{} is scalable and can support high
throughput of transactions (\secref{eval}), rendering weaker consistency models
unnecessary.

\insertfig{0.95}{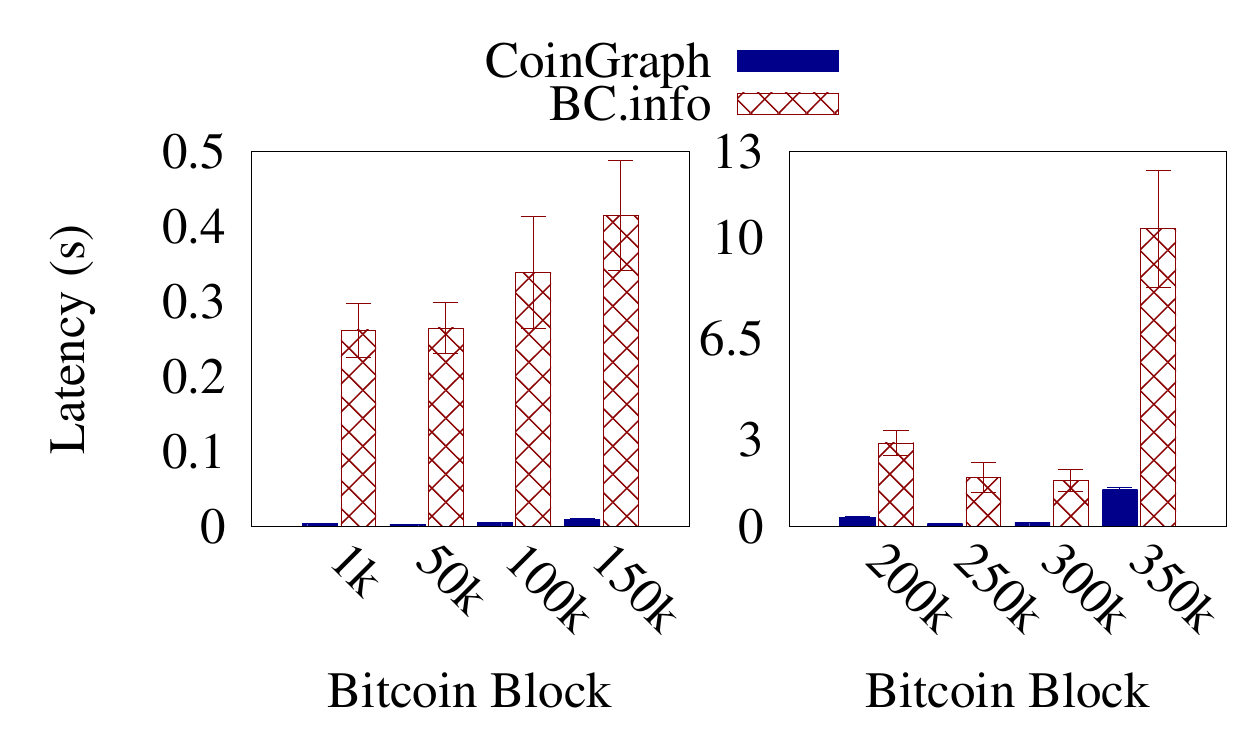}{Average latency (secs) of a
Bitcoin block query in blockchain application.  CoinGraph, backed by
\sysname{}, is an order of magnitude faster than
Blockchain.info.}{blockchain_latency}

\insertfig{0.8}{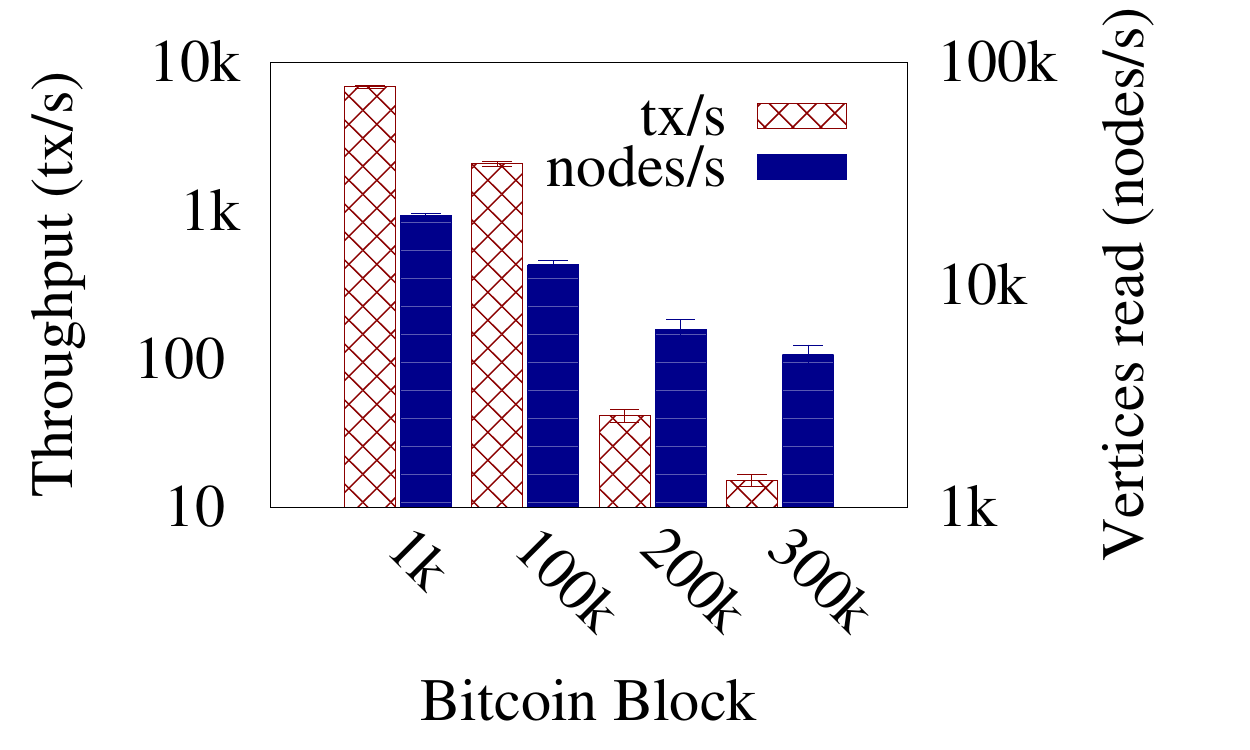}{Throughput of Bitcoin block render queries
in CoinGraph.  Each query is a multi-hop node program.  Throughput decreases as
block size increases since higher blocks have more Bitcoin transactions (more
nodes) per query.}{cg_xput}

\vspace{-0.5\baselineskip}
\section{Evaluation}\label{sec:eval}
In this section, we evaluate the performance of refinable timestamps in a
\sysname{} implementation comprising 40K lines of C++ code.  Our evaluation
shows that \sysname{}:
\begin{itemize}
    \item enables CoinGraph to execute Bitcoin block queries $\bcratio\times$
        faster than Blockchain.info\ct{bcinfo} (\secref{eval_coingraph});
    \item outperforms Titan\ct{titan} by $\titanratiotao\times$ on social
        network workload\ct{tao} (\secref{eval_titan}) and outperforms
        GraphLab\ct{powergraph} by $\graphlabratio\times$ on node program
        workload (\secref{eval_graphlab});
    \item scales linearly with the number of gatekeeper and shard servers for
        graph analysis queries (\secref{eval_scale});
    \item balances the tension between proactive and reactive ordering overheads
        (\secref{eval_tau}).
\end{itemize}

\subsection{CoinGraph}\label{sec:eval_coingraph}
To evaluate the performance of CoinGraph we deploy \sysname{} on a cluster
comprising 44 machines, each of which has two 4 core Intel Xeon 2.5 GHz L5420
processors, 16 GB of DDR2 memory, and between 500 GB and 1 TB SATA spinning
disks from the same era as the CPUs.  The machines are connected with gigabit
ethernet via a single top of rack switch, and each machine has 64-bit Ubuntu
14.04 and the latest version of Weaver and HyperDex Warp\ct{warp}.  The total data
stored by CoinGraph comprises more than 1.2B edges and occupies $\sim 900$ GB on
disk, which exceeds the cumulative memory (704 GB).  We thus implement demand
paging in \sysname{} to read vertices and edges from HyperDex Warp in to the memory
of \sysname{} shards to accommodate the entire blockchain data.

\begin{table}
\begin{tabular}{|r|l|l|}
  \hline
    & \texttt{get\us edges} & $59.4\%$ \\ \cline{2-3}
    \textbf{Reads 99.8\%} & \texttt{count\us edges} & $11.7\%$ \\ \cline{2-3}
    & \texttt{get\us node} & $28.9\%$ \\ \cline{2-3}
  \hline
    & \texttt{create\us edge} & $80.0\%$ \\ \cline{2-3}
    \textbf{Writes 0.2\%} & \texttt{delete\us edge} & $20.0\%$ \\ \cline{2-3}
  \hline
\end{tabular}
\caption{Social network workload based on Facebook's TAO.}\label{tab:tao_workload}
\end{table}

We first examine the latency of single Bitcoin block query, averaged over 20
runs.  A block query is a node program in \sysname{} that starts at the Bitcoin
block vertex, and traverses the edges to read the vertices that represent the
Bitcoin transactions that comprise the block.  We calibrate CoinGraph's
performance by comparing with Blockchain.info\ct{bcinfo}, a state-of-the-art
commercial block explorer service backed by MySQL\ct{bcinfo_arch}.  We use their
blockchain raw data API that returns data identical to CoinGraph in JSON format.


\begin{figure}
\centering
    \begin{subfigure}{.48\linewidth}
        \centering
        \includegraphics[width=0.99\textwidth]{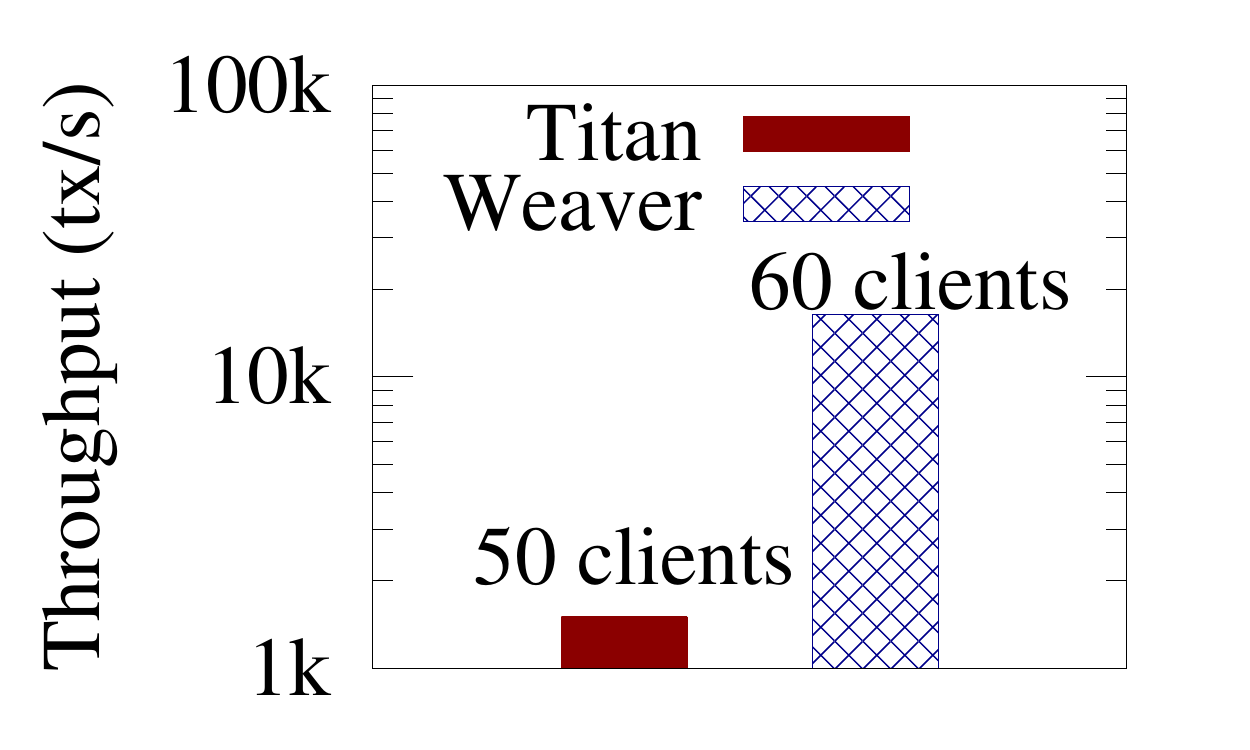}
        \caption{Social Network Workload}
        \label{fig:tao_tput}
    \end{subfigure}
    \begin{subfigure}{.48\linewidth}
        \centering
        \includegraphics[width=0.99\textwidth]{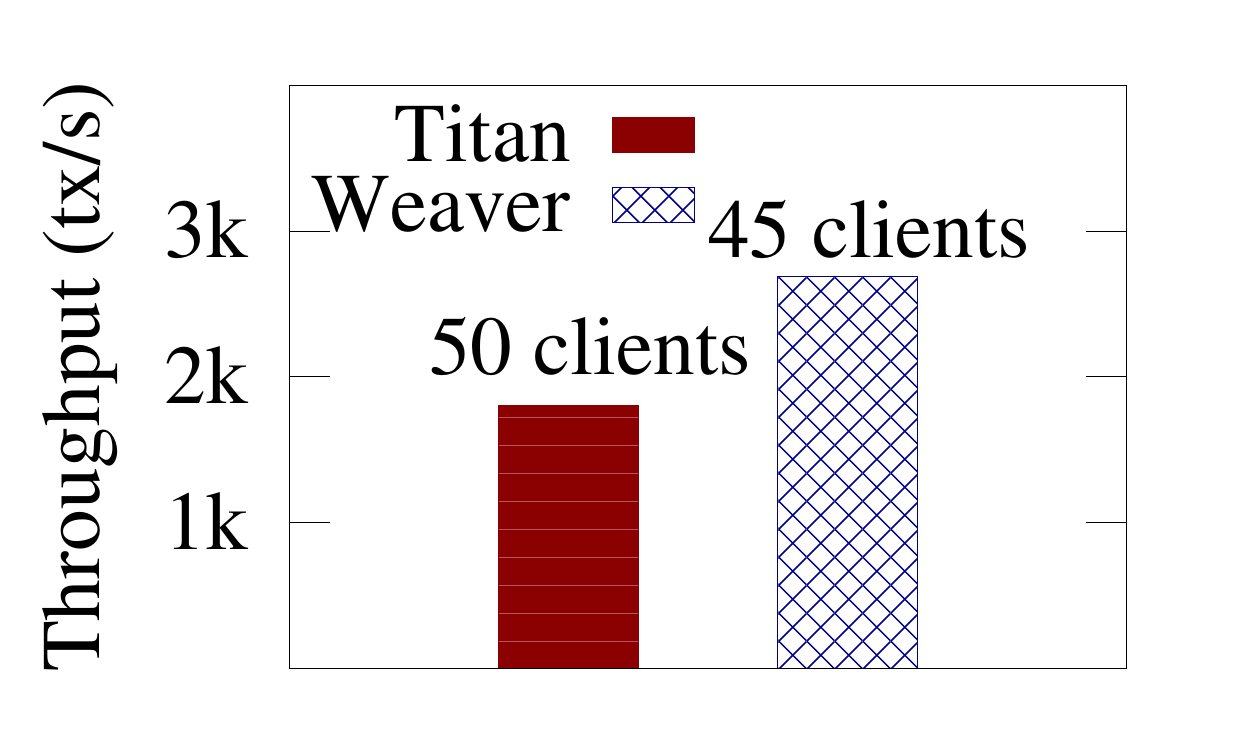}
        \caption{75\% Read Workload}
        \label{fig:75_tput}
    \end{subfigure}
\caption{Throughput on a mix of read and write transactions on the LiveJournal
    graph.  \sysname{} outperforms Titan by $\titanratiotao\times$ on a
    read-heavy TAO workload, and by $\titanratiow\times$ on a 75\% read
    workload.  The numbers over each bar denote the number of concurrent clients
    that issued transactions.  Reactively ordered transactions comprised
    0.0013\% of the TAO workload and 1.7\% of the 75\% read workload.}
\label{fig:eval_tput}
\end{figure}

The results (\figref{blockchain_latency}) show that the performance of block
queries is proportional to the number of Bitcoin transactions in the block for
both systems, but CoinGraph is significantly faster.  Blockchain.info's absolute
numbers in \figref{blockchain_latency} should be interpreted cautiously as they
include overheads such as WAN latency (about 0.013s) and concurrent load from
other web clients.  The critical point to note is that CoinGraph takes about
0.6--0.8ms per transaction per block, whereas Blockchain.info takes 5--8ms per
transaction per block.  The marginal cost of fetching more transactions per
query is an order of magnitude higher for Blockchain.info, due to expensive
MySQL join queries.  Weaver's lightweight node programs enable CoinGraph to
fetch block 350,000, comprising 1795 Bitcoin transactions, $\bcratio\times$
faster than Blockchain.info.


We also evaluate the throughput of block queries supported by CoinGraph.
\figref{cg_xput} reports the variation in throughput of the system as a function
of the block number.  In this figure, the data point corresponding to block $x$
reports the throughput, averaged over multiple runs, of executing block node
programs in CoinGraph for blocks randomly chosen in the range $[x,x+100]$.
Since each node program is reads many vertices, \figref{cg_xput}
also reports the rate of vertices read by the system.  The system is able to
sustain node programs that perform 5,000 to 20,000 node reads per second.

\subsection{Social network benchmark}\label{sec:eval_titan}
We next evaluate \sysname{}'s performance on Facebook's TAO workload\ct{tao}
(\tableref{tao_workload}) using a snapshot of the LiveJournal social
network\ct{livejournal_data} comprising 4.8M nodes and 68.9M edges (1.1 GB).
The workload consists of a mix of reads (node programs in \sysname{}) and writes
(transactions in \sysname{}) that represent the distribution of a real social
network application.  Since the workload consists of simple reads and writes,
this experiment stresses the core transaction ordering mechanism.  We compare
\sysname{}'s performance to Titan\ct{titan}, a graph \store{} similar to
\sysname{} (distributed, OLTP) implemented on top of key-value stores.  We use
Titan v0.4.2 with a Cassandra backend running on identical hardware.  We use a
cluster of 14 machines similar to those in \secref{eval_coingraph}.

\insertfig{0.9}{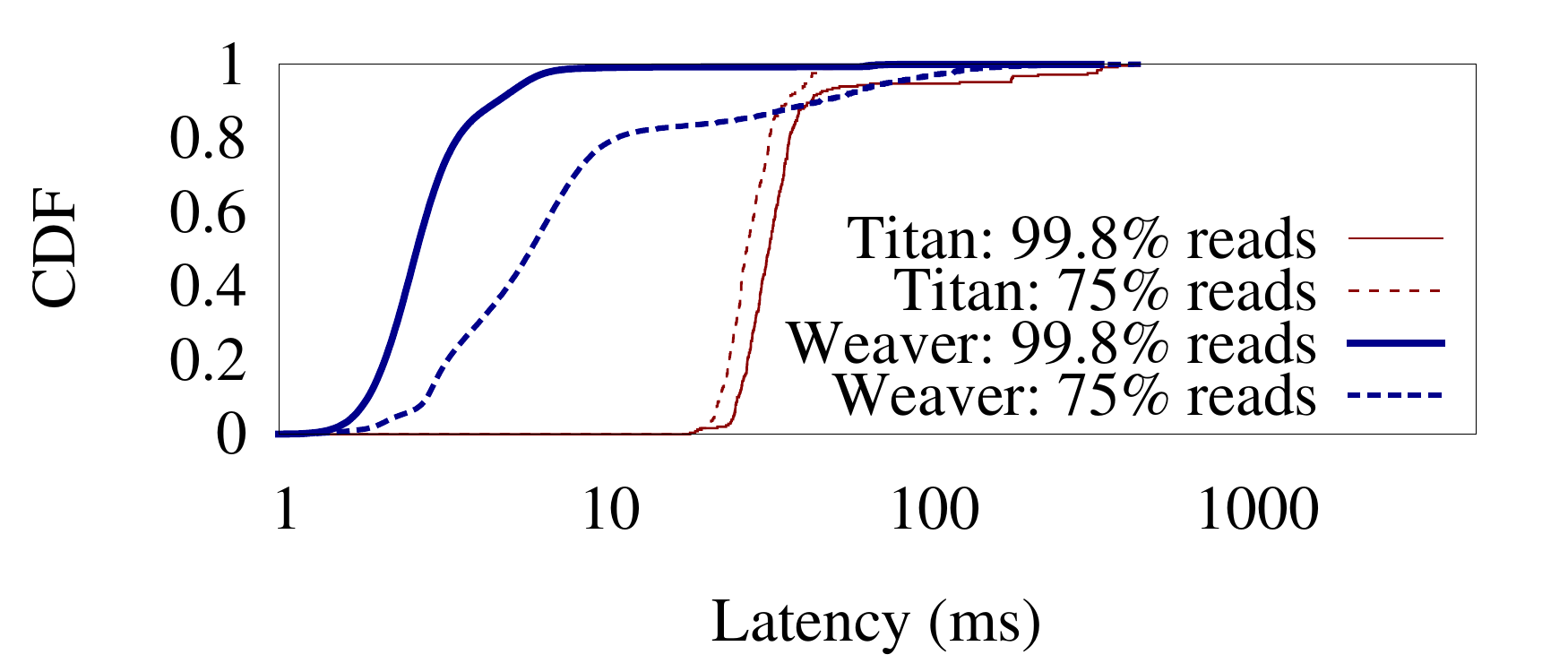}{CDF of transaction latency for a
social network workload on the LiveJournal graph.  \sysname{} provides
significantly lower latency than Titan for all reads and most
writes.}{online_latency}

\insertfig{0.9}{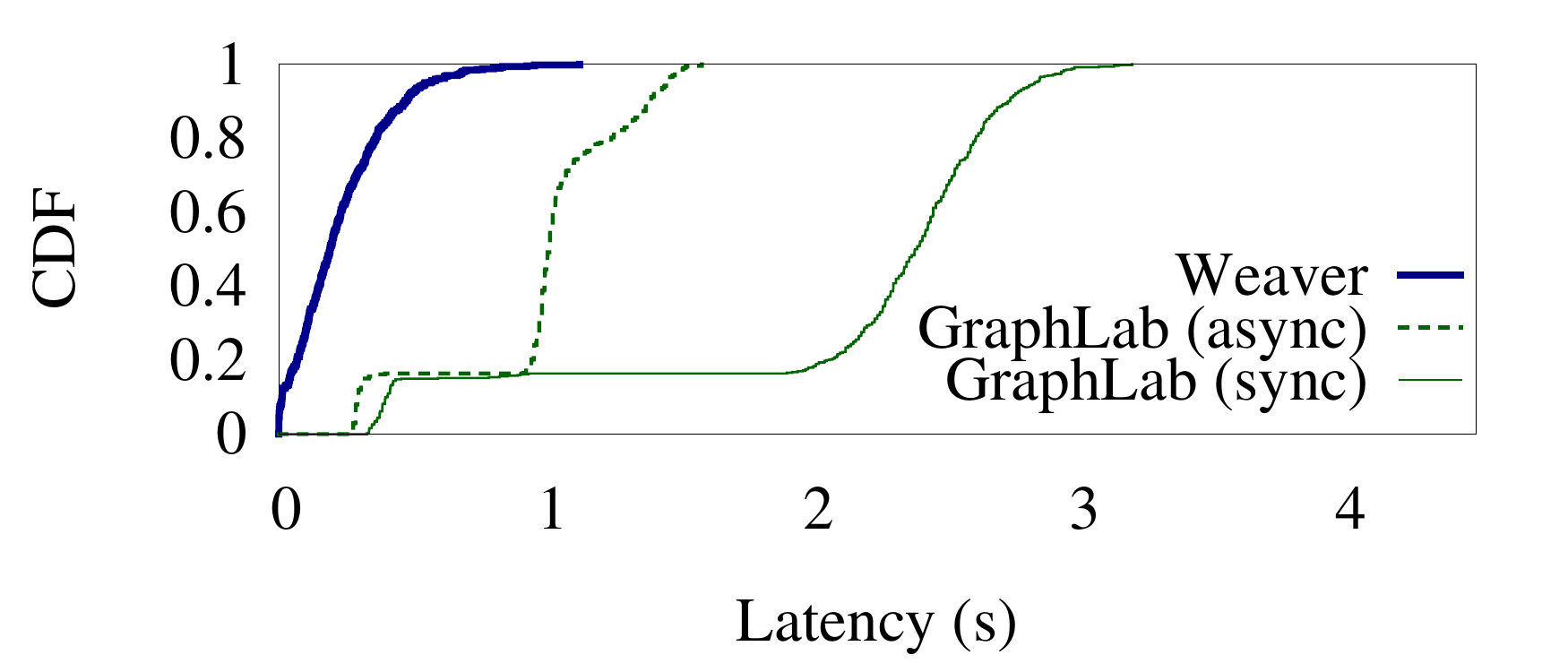}{CDF of latency of traversals on
the small Twitter graph.  \sysname{} provides $4.3\times$-$9.4\times$ lower
latency than GraphLab in spite of supporting mutating graphs with
transactions.}{proc_latency}

\tightpara{Throughput}
\figref{tao_tput} shows the throughput of \sysname{} compared to Titan.
\sysname{} outperforms Titan by a factor of $\titanratiotao\times$.  \sysname{}
also significantly outperforms Titan across benchmarks that comprise different
fractions of reads and writes as shown in \figref{75_tput}.

Titan provides limited throughput because it uses two-phase commit with
distributed locking in the commit phase to ensure
serializability\ct{titan_slides}.  Since it always has to pessimistically lock
all objects in the transaction, irrespective of the ratio of reads and writes,
Titan gives nearly the same throughput of about 2000 transactions per second
across all the workloads.  \sysname{}, on the other hand, executes graph
transactions using refinable timestamps leading to higher throughput for all
workloads.

\sysname{}'s throughput decreases as the percentage of writes increases.  This
is because the timeline oracle serializes concurrent transactions that modify
the same vertex.  \sysname{}'s throughput is higher on read-mostly workloads
because node programs can execute on a snapshot of the graph defined by the
timestamp of the transaction.

\tightpara{Latency}
\figref{online_latency} shows the cumulative distribution of the transaction
latency on the same social network workloads.  We find that node program
execution has lower latency than write transactions in \sysname{} because
writes include a transaction on the backing store.  As the percentage of writes
in the workload increases, the latency for the requests increases.  In contrast,
Titan's heavyweight locking results in higher latency even for reads.

\vspace{-0.5\baselineskip}
\subsection{Graph analysis benchmark}\label{sec:eval_graphlab}
Next, we evaluate \sysname{}'s performance for workloads which involve more
complicated, traversal-oriented graph queries.  Such workloads are common in
applications such as label propagation, connected components, and graph
search\ct{fb_graphsearch}.  For such queries, we compare \sysname{}'s
performance to GraphLab\ct{powergraph} v2.2, a system designed for offline graph
processing.  Unlike \sysname{}, GraphLab can optimize query execution without
concern for concurrent updates.  We use both the synchronous and
asynchronous execution engines of GraphLab.  We use the same cluster as in
\secref{eval_titan}.

The benchmark consists of reachability queries on a small Twitter
graph\ct{twitter_data} consisting of 1.76M edges (43 MB) between vertices chosen
uniformly at random.  We implement the reachability queries as breadth-first
search traversals on both systems.  In order to match the GraphLab execution
model, we execute \sysname{} programs sequentially with a single client.

The results show that that, in spite of supporting strictly serializable online
updates, \sysname{} achieves an average traversal latency that is
$\graphlabratio\times$ lower than asynchronous GraphLab and $9\times$ lower than
synchronous GraphLab.  \figref{proc_latency} shows that the latency variation
for this workload is much higher as compared to the social network workload,
because the amount of work done varies greatly across requests.  Synchronous
GraphLab uses barriers, whereas asynchronous GraphLab prevents neighboring
vertices from executing simultaneously---both these techniques limit concurrency
and adversely affect performance.  \sysname{} allows a higher-level of
concurrency due to refinable timestamps.

\insertfig{0.65}{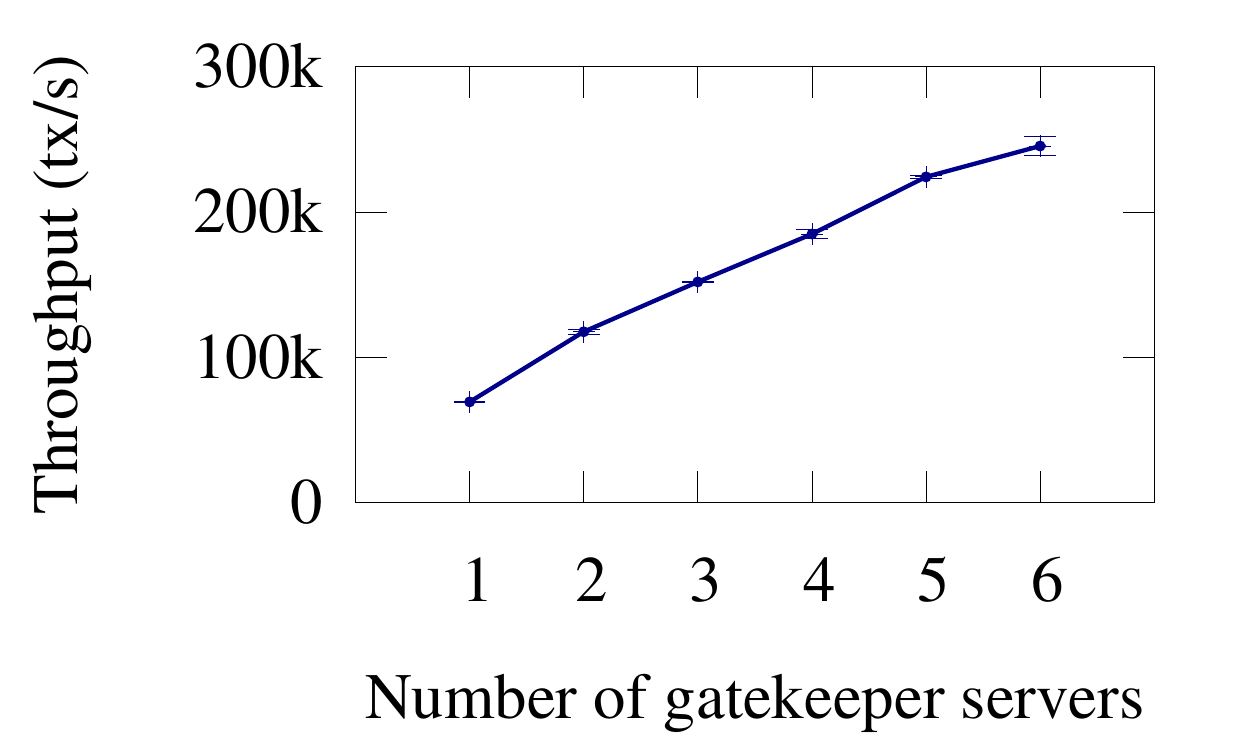}{Throughput of \texttt{get\us
node} programs. \sysname{} scales linearly with the number of gatekeeper
servers.}{vts_scale}

\insertfig{0.65}{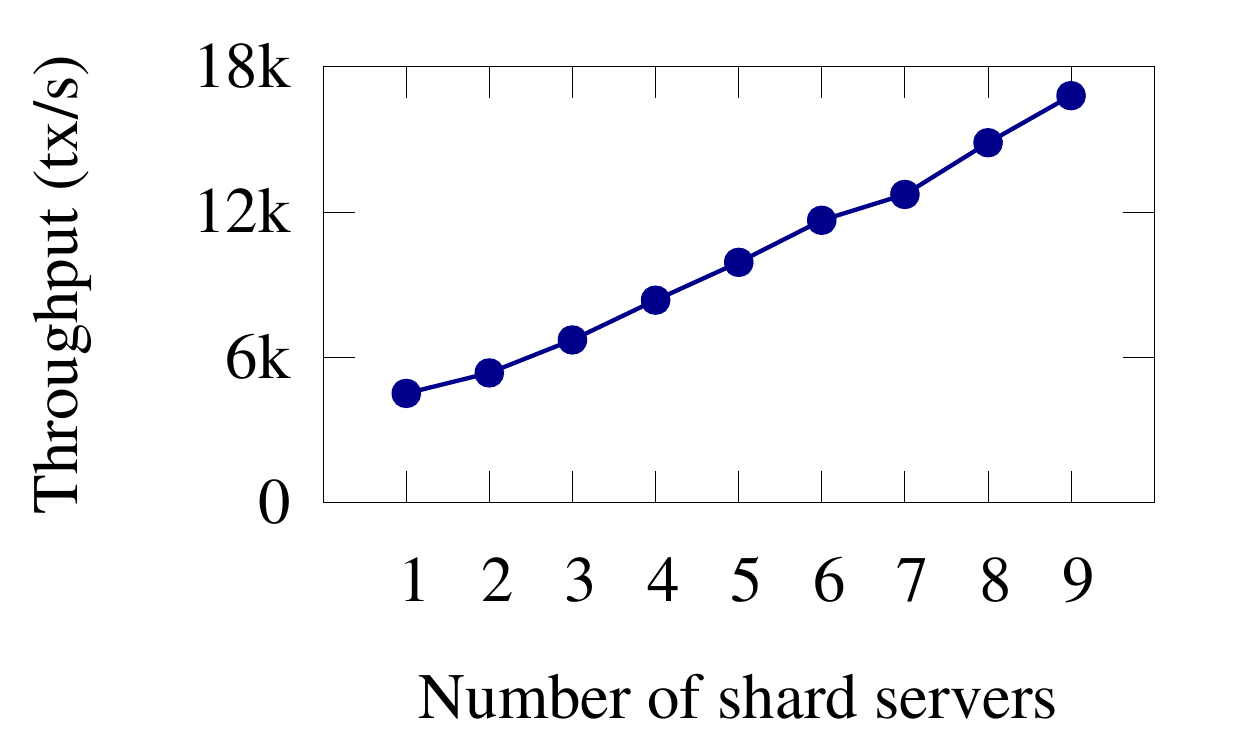}{Throughput of local clustering
coefficient program. \sysname{} scales linearly with the number of shard
servers.}{shard_scale}

\subsection{Scalability}\label{sec:eval_scale}
To investigate how \sysname{}'s implementation of refinable timestamps scales,
we measure \sysname{}'s throughput on microbenchmarks with varying number of
servers.  We perform the first set experiments on an Amazon EC2 cluster
comprising 16 r3.2xlarge instances, each running Ubuntu 14.04 on an 8 core Intel
Xeon E5-2670 (Ivy Bridge) processor, 61 GB of RAM, and 160 GB SSD storage.  We
perform this experiment on the Twitter 2009 snapshot\ct{twitter_rv} comprising
41.7M users and 1.47B links (24.37 GB).

\figref{vts_scale} shows the throughput of \texttt{get\us node} node programs in
\sysname{} with a varying number of gatekeeper servers.  Since these queries are
local to individual vertices, the shard servers do relatively less work and the
gatekeepers comprise the bottleneck in the system.  \sysname{} scales to about
250,000 transactions per second with just 6 gatekeepers.

However, as the complexity of the queries increases, the shard servers perform
more work compared to the gatekeeper.  The second scalability microbenchmark,
performed on a small Twitter graph with 1.76M edges (43 MB) \ct{twitter_data} using the
same cluster as \secref{eval_titan}, measures the performance of the system
on local clustering coefficient node programs.  These programs require more work
at the shards: each vertex needs to contact all of its neighbors, resulting in a
query that fans out to one hop and returns to the original vertex.
\figref{shard_scale} shows that increasing the number of shard servers, while
keeping the number of gatekeepers fixed, results in linear improvement in the
throughput for such queries.

The scalability microbenchmarks demonstrate that \sysname{}'s transaction
ordering mechanism scales well with additional servers, and also describe how
system administrators should allocate additional servers based on the workload
characteristics.  In practice, an application built on \sysname{} can achieve
additional, arbitrary scalability by turning on node program caching
(\secref{partition_cache}) and also by configuring read-only replicas of shard
servers if weaker consistency is acceptable, similar to TAO\ct{tao}.  We do not
evaluate these mechanisms as they are orthogonal to transaction ordering.

\insertfig{0.8}{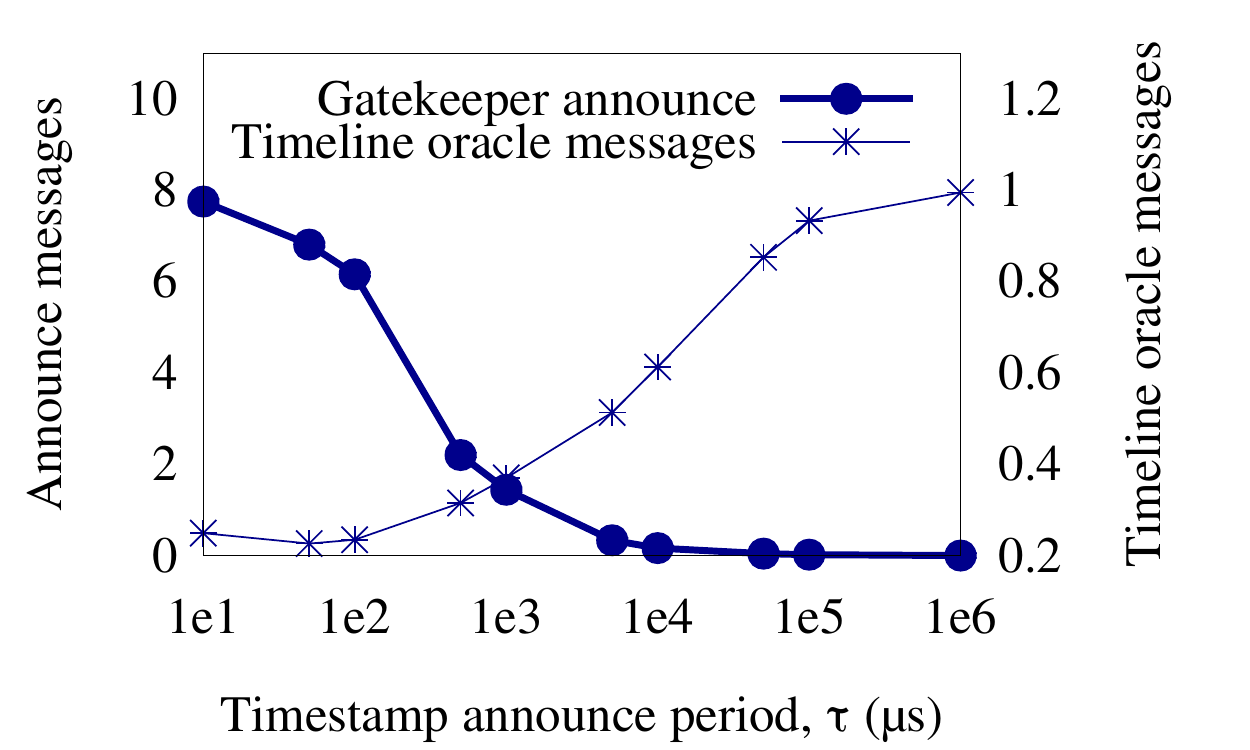}{Coordination overhead, measured in
terms of timestamp announce messages and timeline oracle calls, normalized by
number of queries.  High clock announce frequency results in large gatekeeper
coordination overhead, whereas low frequency causes increased timeline oracle
queries.}{gossip}

\subsection{Coordination Overhead}\label{sec:eval_tau}
Finally, we investigate the tension between proactive (gatekeeper announce
messages) and reactive (timeline oracle queries) coordination in \sysname{}'s
refinable timestamps implementation.  The fraction of transactions which are
ordered proactively versus reactively can be adjusted in \sysname{} by varying
the vector clock synchronization period $\tau$.

To evaluate this tradeoff, we measured the number of coordination messages due
to both gatekeeper announces and timeline oracle queries, as a function of
$\tau$, to order the same number of transactions.  \figref{gossip} shows that
for small values of $\tau$, the vector clocks are sufficient for ordering a
large fraction of the requests.  As $\tau$ increases, the reliance on the
timeline oracle increases.  Both extremes are undesirable and result in high
overhead---low values of $\tau$ waste gatekeeper CPU cycles in processing
announce messages, while high values of $\tau$ cause increased latency to
due extra timeline oracle messages.  An intermediate value represents a good
tradeoff leading to high-throughput timestamping with
occasional concurrent transactions consulting the timeline oracle.

\vspace{-0.5\baselineskip}
\section{Related Work}\label{sec:relwork}
Past work on distributed data storage and graph processing can be roughly
characterized as follows.

\tightpara{Offline Graph Processing Systems} Google's Pregel\ct{pregel}
computation paradigm has spurred a spate of recent
systems\ct{giraph,gps,naiad,powergraph,graphchi,xstream,grace,grace_cornell,laptop_graph,powerlyra,graphx,gridgraph,graphq,pregelix,surfer,hama,pregelplus,blogel,mocgraph,giraphpp,giraphuc,mizan,fb_giraph}
designed for offline processing of large graphs.  Such systems do not support
the rich property graph abstraction, transactional updates, and lookup
operations of a typical graph \store{}.

\tightpara{Online Graph \Store{}s}
The Scalable Hyperlink Store \cite{microsoft_shs} provides the property graph
abstraction over data but does not support arbitrary properties on vertices and
edges.  Trinity\ct{trinity} is a distributed graph \store{} that does not
support ACID transactions.  SQLGraph\ct{sqlgraph} embeds property graphs in a
relational database and executes graph traversals as SQL queries. TAO\ct{tao} is
Facebook's geographically distributed graph backend (\secref{socnet_app}).
Titan\ct{titan} supports updates to the graph and a vertex-local query model.

Centralized graph \store{}s are suitable for a number of graph processing
applications on non-changing, static graphs\ct{laptop_graph}.  However,
centralized \datastore{}s designed for online, dynamic
graphs\ct{neo4j,hypergraphdb,graphchidb,deuteronomy} pose an inevitable
scalability bottleneck in terms of both concurrent query processing and graph
size.  It is difficult to support the scale of modern content
networks\ct{fb_anatomy,tao} on a single machine.

\tightpara{Temporal Graph \Store{}s}
A number of related systems\ct{chronos,kineo,disk_temporal_graphs} are designed
for efficient processing of graphs that change over time.  Chronos\ct{chronos}
optimizes for spatial and temporal locality of graph data similar to \sysname{},
but it does not support ACID transactions.


Kineograph\ct{kineo} decouples updates from queries and executes queries on a
stale snapshot. It executes queries on the last available snapshot of the graph
while new updates are delayed and buffered until the end of 10 second epochs.
In contrast, refinable timestamps enable low-latency updates
(\secref{eval_titan}, \secref{eval_graphlab}) and ensure that node programs
operate on the latest version of the graph.

\tightpara{Consistency Models}
Many existing \datastore{}s support only weak consistency models, such as
eventual consistency\ct{tao,fb_consistency}.
\sysname{} provides the strongest form of data
consistency---strict serializability---as do few other contemporary
systems\ct{spanner,warp,neo4j,titan}.

\tightpara{Concurrency Control}
Pessimistic two-phase locking\ct{2pl} ensures correctness and strong consistency
but excessively limits concurrency.  Optimistic concurrency control
techniques\ct{occ} (OCC) are feasible in scenarios where the expected
contention on objects is low and transaction size is small.  FaRM\ct{farm_sosp}
uses OCC and 2PC with version numbers over RDMA-based messaging.  Graph
\store{}s that support queries that touch a large portion of the graph are not
well-served by OCC techniques.

\sysname{} leverages refinable timestamps to implement multi-version concurrency
control\ct{mvcc,fast_mvcc}, which enables long-running graph algorithms to read
a consistent snapshot of the graph.  Bohm\ct{bohm} is a similar MVCC-based
concurrency control protocol for multi-core settings which serializes timestamp
assignment at a single thread.  Centiman\ct{centiman} introduces the watermark
abstraction---the timestamp of the latest completed transaction---over
traditional logical timestamps or TrueTime.  Weaver uses a similar abstraction
for garbage collection (\secref{gc}) and node programs (\secref{impl_nodeprog}).
Deuteronomy\ct{deuteronomy} is a centralized, multi-core database that
implements MVCC using a latch-free transaction table.

\vspace{-0.5\baselineskip}
\section{Conclusion}\label{sec:concl}

This paper proposed refinable timestamps, a novel, highly scalable mechanism for
achieving strong consistency in a distributed \datastore{}.  The key idea behind
refinable timestamps is to enable a coarse-grained ordering that is sufficient
to resolve the majority of transactions and to fall back on a finer-grained
timeline oracle for concurrent, conflicting transactions.  \sysname{} implements
refinable timestamps to support strictly serializable and fast transactions as
well as graph analyses on dynamic graph data.  The power of refinable timestamps
enables \sysname{} to implement high-performance applications such as CoinGraph
and RoboBrain which execute complicated analyses on online graphs.

\tightpara{Acknowledgments}
We are grateful to the anonymous reviewers as well as colleagues who provided
insightful feedback on earlier drafts of the paper.  This material is based upon
work supported by the National Science Foundation under Grant No. CNS-1518779.
Any opinions, findings, and conclusions or recommendations expressed in this
material are those of the authors and do not necessarily reflect the views of
the National Science Foundation.

\bibliographystyle{abbrv}
\bibliography{../xtx/bib_weaver.xtx}
\end{document}